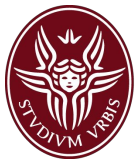

# Towards life-long learning of posture control for s-EMG prostheses

**Facoltà di Ingegneria dell'Informazione, Informatica e Statistica**
**Corso di laurea in Ingegneria dell'Informazione**

Marco Lampacrescia
n° matricola 1530514

Relatore:

Prof.ssa Barbara Caputo

Correlatore:

Prof. Umberto Nanni

A/A 2014/2015

# Table of contents

**Abstract**

Surface electromyography (s-EMG) sensors are a promising way to control upper-limb prostheses. However a training session is necessary in order to set up the controller that will make s-EMG based movement possible.

All data recorded during the training session are used by a machine learning algorithm to make a posture classification, that will allow the controller to distinguish each posture.

The aim of this study is to investigate if it's possible to make a posture classification which can remain valid over time.

The next step will be the study of how it varies depending on the amount of information submitted to it during the training session in view of real life everyday use of the upper-limb prosthesis.

**Introduction**

One of the main difficulties in the rehabilitation of an amputee is the use of prosthetic devices.

Such prostheses actually need non physiological subject's muscle co-contractions. This strategy results in unintuitive usage of these devices, that need significant training period, often considered as stressful or painful in the worst cases [1].

This is why some subjects decide to refuse mechanical prosthesis.

s-EMG sensors are a non invasive way to get a huge set of information about muscle activities. The use of these sensors as control source for upper-limb prostheses has received considerable attention in the last few years, with the aim of making prostheses usage as natural as possible [2].

In order to make it viable the usage of dexterous prosthetic devices, multiple s-EMG sensors are required, so that their output can be combined to effective feature extraction and multidimensional classifier [2].

According to this assertion, firstly we took the raw s-EMG readings using the acquisition protocol shown in Figure 1: it consists of a set of postures proposed by a movie that the subject has to replicate several times [3]. All the postures we studied during this work are represented in Figure 2.

Secondly we performed the features extraction on the raw readings: features extraction consists of a transformation of the raw readings of the s-EMG sensors, in order to allow the classifier algorithm to use them.

Then we used Pattern Recognition approach to extract information from the featured s-EMG data and classify them in a motion class, which allows to distinguish each posture: this step is also known as "training session" [2].

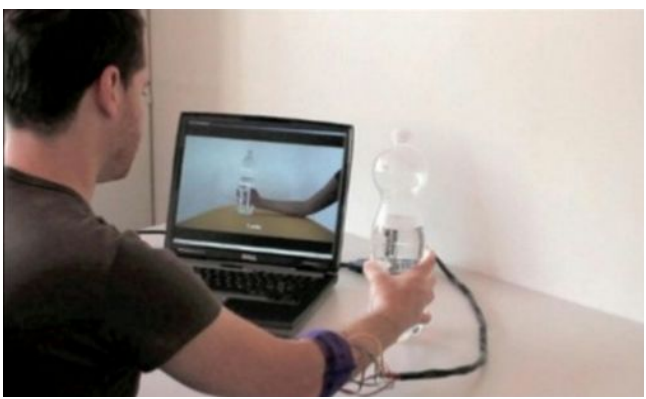

*Fig. 1: Acquisition protocol during the training session.*

*Image courtesy of "HES-SO Haute école spécialisée de Suisse occidentale"*
*at http://www.slideshare.net*

On one hand it is unlikely that this classification remains valid over time because of displacement sensors in usage, user fatigue, and many other factors that make s-EMG sensors data different for each use [7].
On the other hand doing the training session at the beginning of each prosthesis use could become stressful, so it's desirable that classification remains valid as long as possible.

With this study we tried to understand if it's possible to make classification valid over time.
To this purpose we made various classifications with increasing raw readings number, in order to test the stability of those classifications over time.

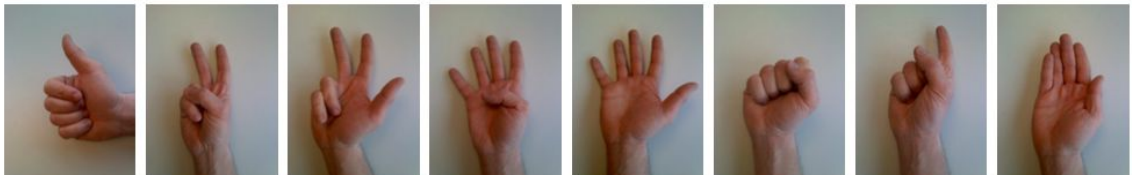

(b) Isometric, isotonic hand configurations ("hand postures").

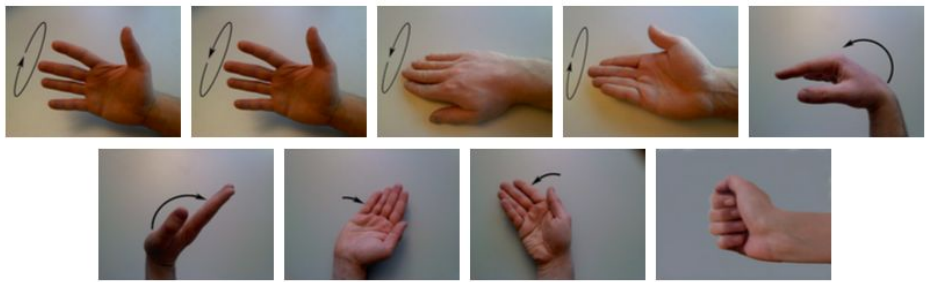

(c) Basic movements of the wrist.

Fig. 2: All postures considered in our article. Image taken from the article [3].

**Related Work**

The control of a mechanical prosthesis through s-EMG sensors is the main argument of a great number of previous works.

The data on which our work is based were made by NinaPro team [4] according to their acquisition protocol [3].

Our pipeline is based on their work and is commonly used in literature. It could be subdivided into these steps: data acquisition, preprocessing, features extraction and final classification [6]. A brief description of the pipeline could be found on Figure 3.

The selection of appropriate features too is based on the results of this article. According to its result, we selected two kind of features:
- Time domain features
- Time-frequency domain features

Typically, time-frequency domain features contain more information about s-EMG signals than time domain features at the cost of increased computation [6].

According to the statement by Zecca *et al.* [5] instant acquisition data don't contain information about the movement that the subject is going to make.

This is why each s-EMG signal is segmented into windows on which various features are then extracted.

Finally we chose only one classifier: Support Vector Machine.

The selection of a single classifier could be supported by the belief that feature representations contribute more significantly to overall performance than classifiers [6].

Another work reported a set of noise factors that make each single use different from the others [7]. One of these factors is the displacement of sensors over the time, due both to everyday life use and the impossibility to wear the prosthesis precisely as it was worn during its last use.

A larger dataset it's indeed necessary for the classifier training in order to make possible to distinguish postures over time [7].

According to this assertion, classifier was trained with increasing datasets, obtaining different classifications on which final tests were performed.

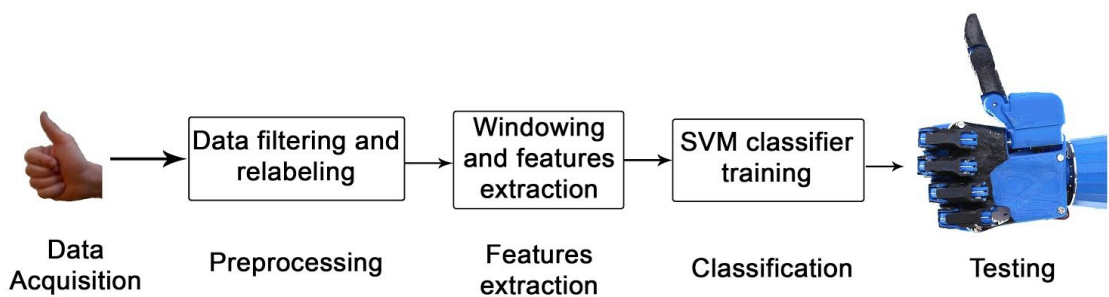

*Fig. 3: A schematic representation of the pipeline of our work.*
  *Mechanical prosthesis image taken from the website: http://www.openhandproject.org*

**Problem Statement**

This work could be subdivided into different steps, from the data acquisition to the classification:
- ➢ data acquisition;
- ➢ preprocessing;
- ➢ feature extraction;
- ➢ classification.

Data acquisition was made by NinaPro team [4], who provided raw data on which this study is based. It consists in four days of acquisitions, with three sessions for each day.

To allow distinction of the various postures all data were recorded by saving the label associated to the posture number shown by the movie mentioned above. Since those data aren't synchronized with the effective movement of the subject and with other noise factors, a preprocessing step is required in order to make reliable test data.

During the preprocessing step s-EMGs and labels data are saved on a single file, then they are interpolated to the highest frequency and at the end the relabeling [6] is performed on them.

Relabeling is a very important thing to achieve good results in the test phase. Thanks to this, synchrony between single movement of the subject and the label associated to it is reliable [6], and the classification step could be more accurate.

After relabeling, we segmented s-EMG signals into windows, in order to perform the features extraction.

Choosing the correct feature representation is an important factor to obtain good accuracy during the test phase.

Their selection was based on results of NinaPro research [6], that indicates Mean Absolute Value (MAV) [6] and Short Time Fourier Transform (STFT) [6] as good feature representation for s-EMG signals.

Results mentioned above indicate Support Vector Machine (SVM) as the classifier that achieved top performance in combination with features mentioned above [6], so the only classifier used was the SVM with Radial Basis Function (RBF) kernel.

We tried different posture classification models obtained by increasing featured datasets, in order to test those classifiers on a test dataset.

To the best of our knowledge there is no prior work investigated the repeatability problem.

**Materials and Methods**
*Data*
Raw data were in '.txt' extension. They are subdivided in two kinds of data:
- emg: it contains s-EMG sensors readings and inclinometer readings. Ten columns are associated to s-EMG sensors and two columns to the inclinometer. The remaining 4 columns are associated with glove measurements, but two of this didn't work, so the values associated to the last two columns are equal to zero.
- movie: it contains the posture label according to the movie shown to the subject during the training session.

In both cases each row contains a timestamp and the information associated to it, as in the example reported on Figure 4.

*Fig. 4: An example of EMG raw data with related timestamps*

Since frequencies of those data are different, labels have to be interpolated to the s-EMG recording frequency (100 Hz).
After this, to remove high frequency equipment noise components, s-EMG data were passed through a low-pass filter as commonly used in the literature.
The final preprocessing step is relabeling [6]. It is necessary, since subjects don't react instantly while watching the video, so the synchronization of labels stored in the movie file with effective movement performed by subject is needed.
The correct time span associated with this movement was found in correspondence of an increased s-EMG activity [6].

*Features Extraction*
As previously said, the choice of a good feature representation is a very important factor to achieve good results during the test phase.
The objective of this is to effectively represent an entire window on single s-EMG signal channel.
We first considered windows of 100, 200 and 400 ms length.
Then only windows with 200 ms length were kept as they lead to better results. During windowing each new window was taken with 10 ms shift from the previous one [6].

Two types of representations have been used on those windows:
- Mean Absolute Value (MAV);
- Short Time Fourier Transform (STFT).

MAV is a Time Domain feature representation. It is computationally easier than STFT and allow posture control through force-related measurements [6].
It consists in a simple mean of absolute measurements contained in a single channel window, so the MAV mathematical definition for each channel's window is as follows:

$$\hat{x} = \frac{1}{T}\sum_{t=1}^{T}|x_t|$$.

The '$T$' value represents the number of acquisitions contained into the entire window.
The '$x_t$' value represents an instant acquisition of a single s-EMG sensor.

As a counterpart, STFT is a Time-Frequency Domain feature representation, so it contains a richer set of information than the MAV representation. In addition it is designed for non stationary signals and it allows better distinction than MAV, at the cost of greater computational complexity.
STFT mathematical definition indicates more than one value for each channel's window:

$$\hat{x}_{k,t} = \sum_{m=0}^{R-1} x_{m-t} g_m e^{-i\frac{2\pi}{M}km}$$.

Here we consider $M$ frequency bins indexed with $k$ [6] and the location in time of the signal indexed with $t$.
We do the transformation by sliding windows obtained through $g$ function of length $R$ [6]. The $g$ function is indexed with the $m$ value, that is the index of the summation and indicates each value of the window function.
The $x$ function represents the acquisitions of a single s-EMG sensors. For its indexing we use both the summation index $m$ and the time index $t$.
After features have been extracted, standardization is performed to have zero mean and unit standard deviation, in order to achieve better accuracy during the classification phase [6].

*Classification:*
The final things to do before testing is posture classification. As mentioned above only one classifier was used: Support Vector Machine (SVM).

*The algorithm:* SVMs are linear binary classifiers that attempt to maximize the margin between the two classes [9] through training samples. Because many problems have more than two classes, many approach could be used to combine multiple two-class SVMs.
One example is the one-versus-the-rest approach, that uses K SVMs for K classes: the k[th] model $y_k(x)$ is trained using the data from class $C_k$ as the positive examples and the data from the remaining K − 1 classes as the negative ones [10].
The key strength of SVMs is the possibility to use kernel function, that allows to use SVMs on nonlinear problems.
The kernel selected for our purpose is the Radial Basis Function (RBF) one, also known as Gaussian kernel. An example of how a kernel works is reported in Figure 5.

*The parameters:* SVM classifier in combination with RBF kernel needs two parameters tuning for its correct usage. Those parameters are known as "C" and "$\gamma$".
"C" is a SVM parameter that trades off misclassification of training examples against simplicity of the decision surface [11].
"$\gamma$" is a RBF kernel parameter that defines how far the influence of a single training example reaches. [11]
The choice of the best "C" and "$\gamma$" parameters has been made using a grid-search approach.

Using this approach, the parameters $C \in \{2^i : i \in \{0, 2, ..., 16\}\}$ and $\gamma \in \{2^i : i \in \{-16, -14, ..., -2\}\}$ have been considered [6], so all possible combinations had been tried, choosing finally the combination that achieves the best validation accuracy. Validation consists in testing the classification model obtained with a single C and $\gamma$ combination with the validation set (more about that later). If the accuracy obtained is better than the previous best one, it's selected as the new best one, otherwise it's discarded.

*The dataset:* In order to properly classify all featured data, we split them in three sections: train, test and validation.
- The 'train' section is the part of dataset used to train the classifier. The choice of a good training set is crucial to obtain reliable classifiers able to achieve good performance during the test phase. The training set has been obtained by taking half of the entire dataset and then reducing it by keeping every tenth sample [6].
- The 'validation' set is used together with the train section on the choice of the best parameters for the classifier training through a grid-search approach. It is composed by random selections of 20% of the remaining half of the dataset.
- The 'test' set is used during the test phase to try out posture classification models previously obtained. It is composed by the remaining part of the dataset.

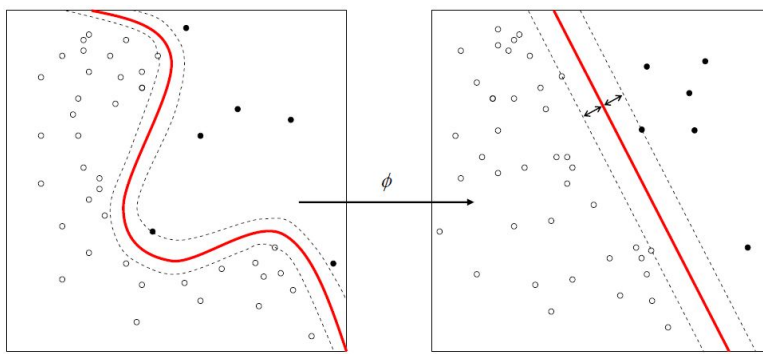

*Fig. 5: A visual representation on how kernel functions work.*
*Image courtesy of Wikipedia at https://en.wikipedia.org/wiki/Kernel_method*

# Experiments

*Experimental setup:*

As mentioned above, to obtain a posture classification that remains valid over time it's needed a large dataset. This is why each training set considered is bigger than the previous one.

For the training dataset, it has been considered the three datasets obtained during first day of acquisitions: respectively the 2nd, the 3rd and the 5th acquisition. So the smallest training dataset is composed only by the 2nd acquisition dataset.

The next one is composed by assembling the 2nd and the 3rd dataset. The last one is composed by assembling the 2nd, the 3rd and the 5th dataset, that means the daily acquisitions.

Each training dataset contains the training set and the validation set, to permit cross validation approach mentioned above. After this step, six classification models had been obtained (three models for each feature type: MAV and STFT). Each posture classification model has to be tested on other days acquisitions. For this scope each first acquisition of the remaining three days was considered: the 8th, the 11th and the 14th one.

In order to increase accuracies we tried to smooth all predicted labels.

This was made by taking the mode value in a range around the predicted label.

Smoothed values have been saved separately from the original one and were used only for accuracy results comparison.

*Results:*

The tables below contains all the validation accuracies reached during the training phase (Table 1), and all accuracy results obtained during the test phase, including smoothing ones (Table 2).

As expected STFT features results in better accuracy performances over MAV ones, that despite of its low computational cost achieves good performances.

As shown in the Table 2, smoothing often permits low performances increasing (on the average of 1% or 2%), but in some cases it decreases label prediction accuracy.

Other information can be extracted from histograms below, that show accuracy on single postures predictions (Figure 9, Figure 10) and overall best (Figure 6), average (Figure 8) and worst (Figure 7) prediction accuracies.

Last result graphs are the confusion matrixes (Figure 11), that clearly illustrate classifiers performances and give information about single postures misclassification and postures that can be evaluated with the best (or the worst) accuracy.

|  | **MAV** | **STFT** |
|---|---|---|
| **Only 2nd training session** | 88.62% | 87.37% |
| **2nd and 3rd tr. sess.** | 87.09% | 84.80% |
| **2nd, 3rd and 5th tr. sess.** | 84.69% | 86.43% |

*Table 1: Validation accuracies of the training phase*

| Train n°: Test n° | MAV Not Sm. | MAV Smooth. | STFT Not Sm. | STFT Smooth. |
|---|---|---|---|---|
| 2: 8 | 72.110 | 72.113 | 77.582 | 79.025 |
| 2: 11 | 71.732 | 73.690 | 73.795 | 74.920 |
| 2: 14 | 74.162 | 76.298 | 74.088 | 76.503 |
| 2, 3: 8 | 76.035 | 76.054 | 79.518 | 79.391 |
| 2, 3: 11 | 75.214 | 77.328 | 76.056 | 76.628 |
| 2, 3: 14 | 75.705 | 76.359 | 75.534 | 77.782 |
| 2, 3, 5: 8 | 75.663 | 77.221 | 82.043 | 83.266 |
| 2, 3, 5: 11 | 73.647 | 75.381 | 77.896 | 78.820 |
| 2, 3, 5: 14 | 74.901 | 75.506 | 78.209 | 80.539 |

*Table 2: All accuracies of the test phase: it contains both not smoothed and smoothed accuracies.*

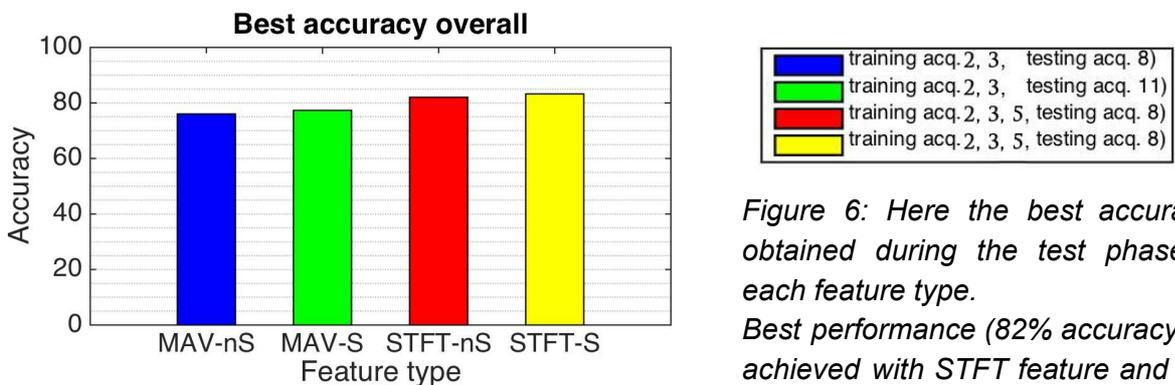

*Figure 6: Here the best accuracies obtained during the test phase for each feature type.*
*Best performance (82% accuracy) are achieved with STFT feature and daily training dataset.*

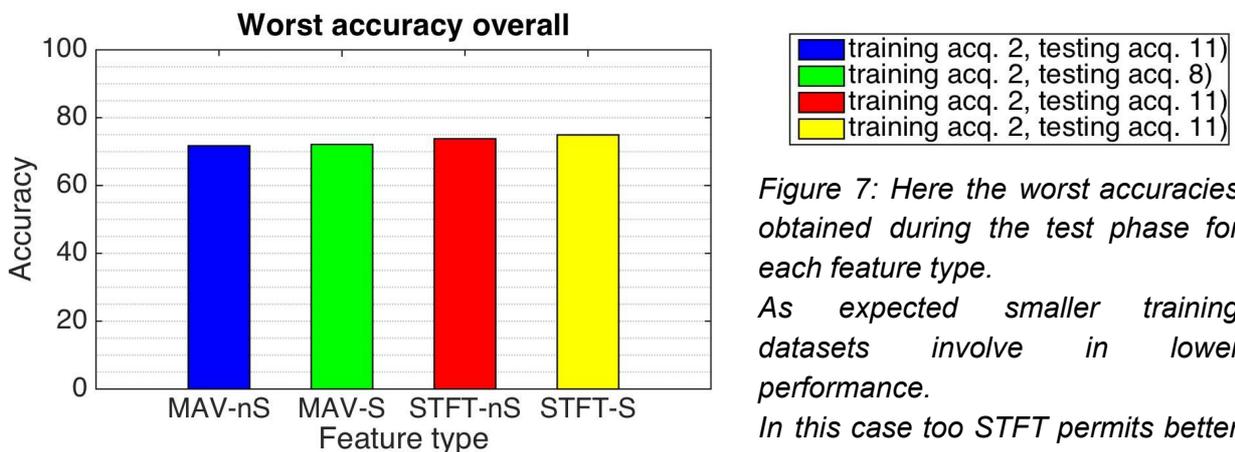

*Figure 7: Here the worst accuracies obtained during the test phase for each feature type.*
*As expected smaller training datasets involve in lower performance.*
*In this case too STFT permits better accuracy.*

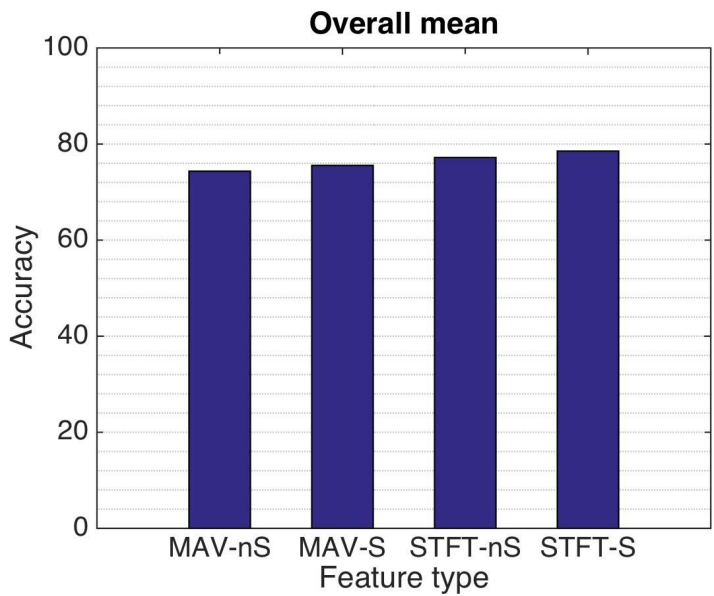

*Figure 8: Here average accuracies for each feature type. As usual STFT permits better accuracy than MAV.*

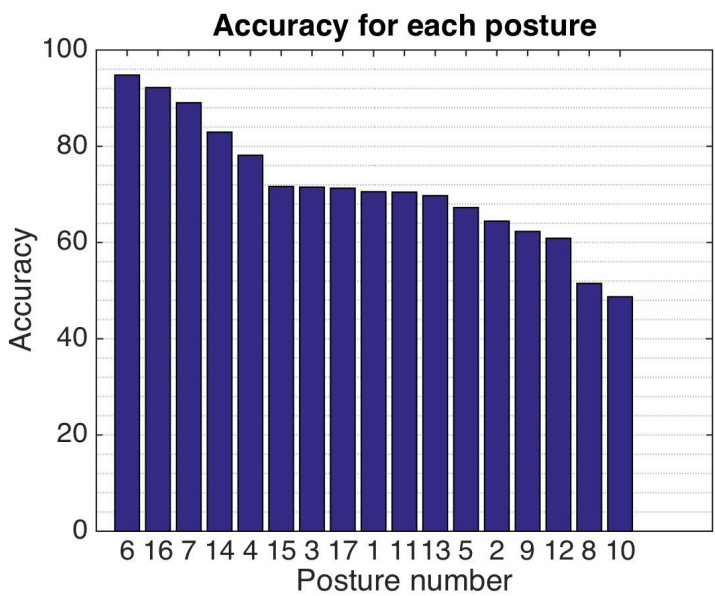

*Figure 9: Here single postures accuracies for the best performing test: training on 2nd, 3rd and 5th dataset and testing on the 8th dataset using STFT features extraction.*

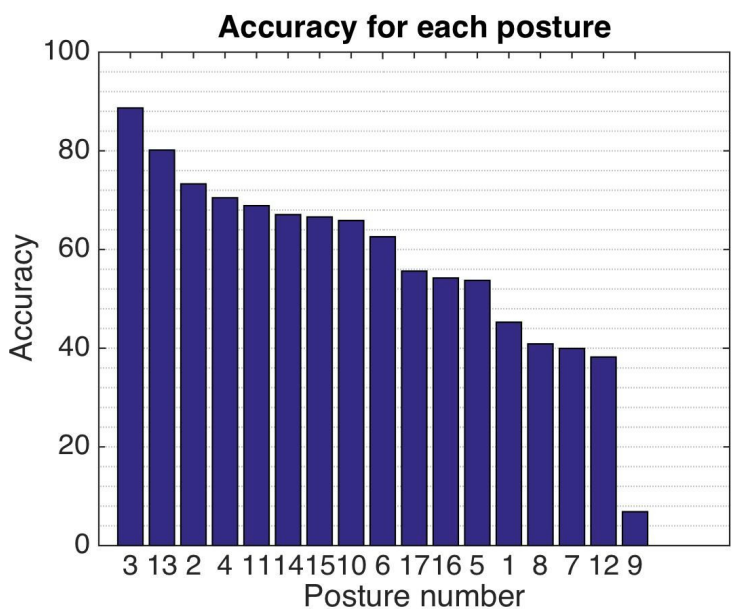

*Figure 10: Here single postures accuracies for the worst performing test: training only on the 2nd dataset and testing on the 11th dataset using MAV features extraction.*
12

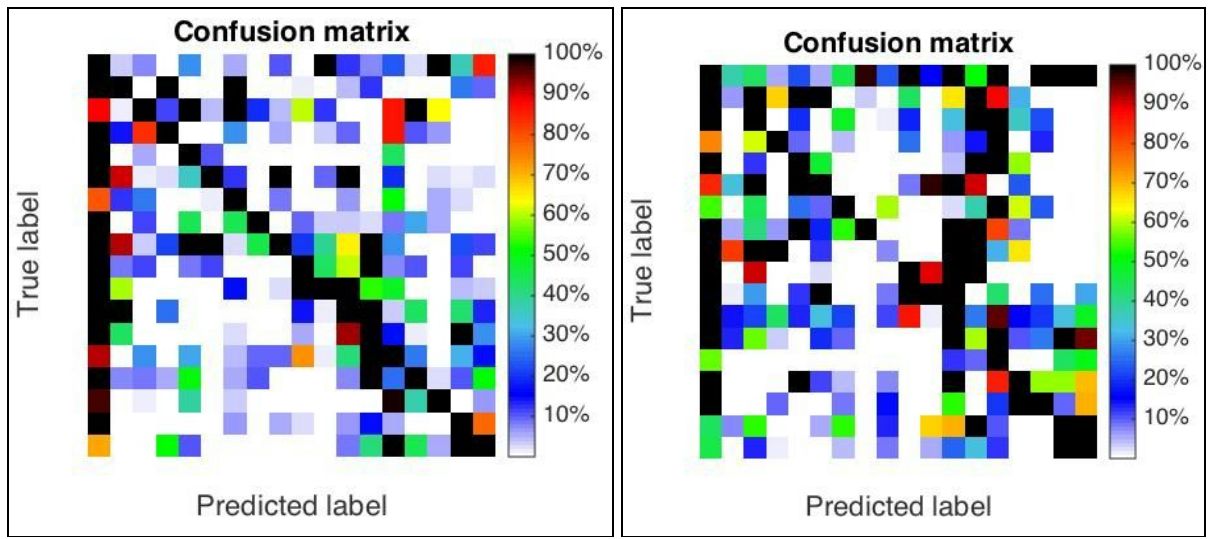

*Figure 11: The image above represent the confusion matrixes relatively to the best performing test (on the left) and the worst one (on the right).*

## Discussion

It's possible to do some consideration on the basis of results reported in the above tables and charts.

The first distinction is between STFT and MAV features. Of course STFT features are better than MAV on posture classification, thanks to STFT properties reported in the previous section. So its usage requires more time than MAV because of its complex mathematical definition and its large set of information. Obviously it's impossible to reach the same result between them, but MAV could be a good proxies for the quality of the datasets.

The main obstacle on making posture classifications valid over time it's a set of variations that make each session of s-EMG sensors use different from the others.

These variations could be caused by electrode conductivity changes, electrophysiological changes, electrode displacement changes, subject cognitive changes and other factors [7].

This is why a single training session is not enough in order to make a classification model able to remain valid over time.

A confirmation for this statement could be found on the "worst accuracy" chart above (Figure 7): all results reported on that are associated to the single dataset training session, independently of the feature type used.

The accuracy of the classifier obtained during the testing phase increase proportionally to training dataset's dimension using STFT features, according to assertions reported in W. Sensinger's article [7], that proposes a dataset obtained by collecting data from areas covering the displacement range, in order to achieve more robust posture classification.

Using MAV features the first increment on dataset results in better accuracy, but after the last increment results became worse than before.

A possible cause could be find in overfitting: it's possible that too big training set involves in a too accurate posture classification, obtaining worse performance during test phase.

On confusion matrixes (Figure 11) it is reported the misclassification results with respect to individual movements.

In both cases (best and worst results), the diagonal components report a good prediction accuracy. This result can also be seen on charts of the single postures (Figure 9, Figure 10), that report consistent results on postures prediction (up to 94% in the best case).

On the other hand confusion matrixes reported that all postures are often misclassified as rest. This might occur for several reasons: a single window can include both rest and non rest samples. Those windows could be tagged with a movement label, even if it contains great portion of rest movement and during the test phase could be classified as rest movement [6]. This problem is introduced by label classification: each label classifies segments or as rest or as a posture and doesn't consider rest to movement segments, that are often misclassified. Relabeling, as explained in the previous section, reduces the amount of mislabeled samples, but it cannot resolve the problem of ambiguous labels mentioned above [6].

Lastly it's uncommon that accuracy reached during the testing phase is greater than accuracy obtained during the training session. This is why average accuracy obtained using STFT features (~80%, see Table 2) could be considered very good against the accuracy reached during training session (~86%, see Table 1) that could be considered as an upper bound.

**Conclusions:**

In this work single subject learning was tested in order to achieve personal postures classifier able to remain valid over time.

In this way, increasing training datasets have been considered with two different feature types.

Experiments resulted in stable accuracy results over time, especially with STFT features combined to daily training, where accuracy remains on an average of 80%. Lower accuracy resulted on MAV features, where accuracy remains stable on an average of 74%.

It should be noted that those experiments were executed on a short time frame (4 days), so its stability over time is unknown. Future work should validate these results on a larger data collection. Moreover, misclassification problems have to be solved in order to achieve reliable datasets on which classifier could be trained.

Research on s-EMG based posture classification is only at the beginning, and those results are far from being considered usable in real-life setting: a prosthesis that misses the correct posture one time out of five may be dangerous for those people who want to use it.

More research have to be done in this direction, in order to make prosthesis as intuitive as possible during its usage.

**Ringraziamenti:**

Ci sarebbe un lungo elenco di persone da ringraziare per questi tre anni di università e per questo importante traguardo che ho appena raggiunto.

Prima di tutti però vorrei ringraziare la prof.ssa Barbara Caputo per tutto il supporto che mi ha dato durante le varie fasi della ricerca e durante la stesura della tesi.

Per quanto riguarda i vari suggerimenti durante lo svolgimento della ricerca vorrei invece ringraziare Ilja Kuzborskij, che ha pazientemente risposto a tutte le mail con le mille domande che gli sono state poste!

Vorrei poi fare un ringraziamento speciale a mio padre, mia madre e mio fratello, che nel bene e nel male, mi hanno sopportato ogni giorno in questi 3 anni di università... e non solo!

Inoltre vorrei ringraziare Francesca Giordaniello e Mara Graziani con cui ho condiviso tutto (o quasi) questo percorso universitario (se ho ancora i capelli, probabilmente, gran parte del merito va a loro).

A seguire ringrazio Dario, Luca e Guido, i miei amici più stretti, grazie ai quali sono riuscito ad arrivare fino a questo momento.

Voglio poi ringraziare altre due persone:

Fabio Sinnona, che con la sua esperienza mi ha permesso un'adeguata preparazione psicologica a tutti i possibili ostacoli che avrei potuto incontrare verso la fine di questi tre anni e con cui ho preparato più di un esame;

Pietro Fiondella, con cui ho iniziato questo percorso universitario e grazie al quale ho avuto modo di conoscere tante altre persone all'interno della facoltà… rendendo la vita da universitario molto più interessante!

Infine voglio ringraziare tutti quelli che ho conosciuto in questi 3 anni, con cui ho condiviso momenti di studio e momenti di svago, e che hanno contribuito a rendere questi miei 3 anni di università uno dei periodi migliori della mia vita.